\documentclass[12pt]{article}
\usepackage[latin1]{inputenc}
\usepackage{amsmath,amssymb}
\usepackage{cite}
\usepackage{times}
\usepackage{mathrsfs} 
\usepackage{siunitx}
\usepackage{calc}
\usepackage{graphicx}
\usepackage{dsfont}
\usepackage{color}
\usepackage{tabularx}
\usepackage{ifthen}
\usepackage{fancyhdr}
\usepackage{appendix}
\usepackage{seqsplit}
\usepackage{caption}
\usepackage{etoolbox}
\usepackage{hyperref}
\apptocmd{\UrlBreaks}{\do\/\do\-}{}{}
\usepackage[capitalise]{cleveref}
\usepackage{placeins}

\newcommand{\rt}{(\vec{r},t)}
\newcommand{\dif}{\mathrm{d}}%

\newcommand{\Nabla}{\vec{\nabla}}%
\newcommand{\ZT}[1]{\textquotedblleft#1\textquotedblright}%
 
\newcolumntype{Y}{>{\centering\arraybackslash}X}%
\newcolumntype{Z}{>{\raggedright\arraybackslash}X}%

\newlength{\myl}%
\newcommand{\SUM}[2]{{\setlength{\myl}{\widthof{$\displaystyle\sum_{#1}^{#2}$}*\real{0.5}-\widthof{$\displaystyle\sum$}*\real{0.5}}\sum_{#1}^{#2}\;\hspace{-\the\myl}}}
\newcommand{\INT}[3]{\settowidth{\myl}{$\displaystyle\int_{#1}^{#2}$}{\int_{#1}^{#2}\;\;\;\hspace{-\the\myl}\dif #3}\,}
\newcommand{\TINT}[3]{\settowidth{\myl}{$\int_{#1}^{#2}$}{\int_{#1}^{#2}\!\ifthenelse{\equal{#1#2}{}}{}{\;\;\;\;\hspace{-\the\myl}}\dif #3}\,}%
\newcommand{\EINT}[3]{\settowidth{\myl}{$\int_{#1}^{#2}$}{\int_{#1}^{#2}\;\;\;\,\hspace{-\the\myl}\dif #3}\,}

\title{\textbf{Containing a pandemic: Nonpharmaceutical interventions and the \ZT{second wave}}} 

\author{Michael te Vrugt,$^{1}$ Jens Bickmann,$^{1}$ Raphael Wittkowski$^{1,\ast}$\\\\
\normalsize{$^{1}$Institut f\"ur Theoretische Physik, Center for Soft Nanoscience,}\\
\normalsize{Westf\"alische Wilhelms-Universit\"at M\"unster, D-48149 M\"unster, Germany}\\\\
\normalsize{$^\ast$Corresponding author; E-mail: raphael.wittkowski@uni-muenster.de}}

\date{}

\begin{document} 
\captionsetup[figure]{labelfont={bf},name={Fig.},labelsep=period}
\maketitle 

\begin{quote}
In response to the worldwide outbreak of the coronavirus disease COVID-19, a variety of nonpharmaceutical interventions such as face masks and social distancing have been implemented. A careful assessment of the effects of such containment strategies is required to avoid exceeding social and economical costs as well as a dangerous \ZT{second wave} of the pandemic. In this work, we combine a recently developed dynamical density functional theory model and an extended SIRD model with hysteresis to study effects of various measures and strategies using realistic parameters. Depending on intervention thresholds, a variety of phases with different numbers of shutdowns and deaths are found. Spatiotemporal simulations provide further insights into the dynamics of a second wave. Our results are of crucial importance for public health policy.
\end{quote}

\clearpage
\section*{Introduction}
The rapid spread of the coronavirus disease 2019 (COVID-19), caused by the severe acute respiratory syndrome coronavirus 2 (SARS-CoV-2) \cite{WuEtAl2020,ZhouEtAl2020,DaviesEtAl2020}, has led governments across the globe to impose severe restrictions on social life, typically denoted \ZT{shutdown} or \ZT{lockdown}. While these have been found to be very effective in reducing the number of infections, they have also been accompanied by high social and economical costs. Moreover, it can be expected that infection numbers rise again after the shutdown has ended (\ZT{second wave}). Consequently, the development of an effective containment strategy that avoids a collapse of both the economy and the healthcare system and that takes into account the problem of multiple outbreaks is of immense public interest.

For this reason, a significant amount of research is currently performed on the effects of various nonpharmaceutical interventions (NPIs) \cite{SaljeEtAl2020,FergusonEtAl2020,DehningZSWNWP2020,MaierB2020,DaviesEtAl2020,RuktanonchaiEtAl2020} and intervention strategies \cite{DaviesEtAl2020,KruseS2020,RuktanonchaiEtAl2020} on the spread of infectious diseases. From a political perspective, the costs associated with different containment measures make it necessary to obtain a detailed understanding of the benefits of various strategies, the effectiveness of different combinations of NPIs, and of whether one type of intervention can compensate for another one. Of particular importance is the question at which stage social restrictions should be imposed and lifted in order to avoid multiple outbreaks and a large number of deaths.

A useful theory for such investigations is the susceptible-infected-recovered (SIR) model developed by Kermack and McKendrick \cite{KermackM1927}, which has been generalized in a large variety of ways in order to incorporate, e.g., governmental interventions \cite{ChladnaKRR2020}. Recently \cite{teVrugtBW2020}, we have proposed an extension of the SIR model based on dynamical density functional theory (DDFT) \cite{Evans1979,MarconiT1999,ArcherE2004,teVrugtLW2020} that incorporates social distancing in the form of a repulsive interaction potential. It allows to treat different types of NPIs separately and therefore provides more detailed insights into containment strategies than the simple SIR model while being computationally more efficient than individual-based models. A further recent development in SIR theory is the description of adaptive containment strategies, which are relevant for the current pandemic \cite{FergusonEtAl2020}, using hysteresis loops \cite{ChladnaKRR2020,KopfovaNRR2020,PimenovKKAOPR2012}.

In this work, we use the SIR-DDFT model and an extended susceptible-infected-recovered-dead (SIRD) model with hysteresis to investigate the effects of various containment strategies with model parameters adapted to the current COVID-19 outbreak in Germany. We compare the effects of face masks and social distancing/isolation and of various threshold values (of the number of infected persons) for imposing and lifting restrictions. Our simulations reveal the existence of various phases with different numbers of outbreaks. This effect needs to be taken into account when making political decisions on shutdown thresholds, as it can significantly affect both the total length of the shutdowns and the number of deaths. Moreover, we show that a second wave can also arise if only one type of restriction is lifted. Finally, it is found that second waves tend to have a different spatial distribution than first waves, an effect that is a current public health concern \cite{Drosten2020}. Our results thereby extend the work in Refs.\ \cite{DehningZSWNWP2020,MaierB2020,KruseS2020,RuktanonchaiEtAl2020}, as they are based on methods from statistical mechanics that allow for deeper insights. Moreover, we break new ground in soft matter physics by developing a DDFT model with time- and history-dependent interaction potential, leading to interesting novel dynamical behavior. This model allows to test a large variety of shutdown strategies and their consequences for the \ZT{second wave} using our freely available code \cite{CodeAndData}.

\section*{\label{face}Nonpharmaceutical interventions: Face masks vs social distancing}
The most widely used theory for modeling disease outbreaks is the SIR model \cite{KermackM1927}. It assumes that the population consists of three groups, namely susceptible (S), infected (I), and recovered (R) individuals. Susceptible persons are infected at a rate $c_{\mathrm{eff}}\bar{I}$, where $c_\mathrm{eff}$ is the effective transmission rate \cite{ChowellHCFH2004}. Infected persons recover at a rate $w$. Recovered persons are immune to the disease. An extension is the SIRD model \cite{BergeLMMK2017}, in which infected persons die at a rate $m$. The governing equations of the SIRD model are
\begin{align}
\dot{\bar{S}} &= - c_{\mathrm{eff}}\bar{S}\bar{I}\label{s},\\  
\dot{\bar{I}} &= c_{\mathrm{eff}}\bar{S}\bar{I} - w\bar{I} - m \bar{I}\label{i},\\ 
\dot{\bar{R}} &= w\bar{I}\label{r}.
\end{align}
We use overbars to distinguish, e.g., the total number of infected persons ($\bar{I}$) from the number of infected persons per unit area ($I$). Due to its simplicity, the SIR(D) model has become very popular and is used in modeling the current coronavirus outbreak, incorporating real data \cite{MaierB2020}. At present, it is not clear whether persons that have recovered from COVID-19 are immune against it, but experiments on rhesus macaques have found that a SARS-CoV-2 infection induces protective immunity against rechallenge \cite{ChandrashekarEtAl2020}.

A drawback of the standard SIR(D) model is the fact that it does not include spatiotemporal dynamics. Moreover, it does not allow to treat various types of NPIs, such as face masks and social distancing, separately. This is possible in individual-based models, which, however, are computationally very expensive. Therefore, an intermediate approach that combines the simplicity of the simple SIR model with the flexibility of individual-based models is very promising in this context. Such an approach is given by the SIR-DDFT model developed in Ref.\ \cite{teVrugtBW2020}. It describes the densities $S$, $I$, and $R$ of susceptible, infected, and recovered persons, respectively, as fields on spacetime governed by the equations
\begin{align}
\partial_tS &= \Gamma_S\Nabla\cdot\bigg( S\Nabla\frac{\delta F}{\delta S}\bigg) -cSI\label{sr},\\  
\partial_tI &= \Gamma_I\Nabla\cdot\bigg( I \Nabla\frac{\delta F}{\delta I}\bigg) +cSI -wI -mI\label{ir},\\ 
\partial_tR &=\Gamma_R\Nabla\cdot\bigg( R \Nabla\frac{\delta F}{\delta R}\bigg) +wI\label{rr}
\end{align}
with time $t$, mobility $\Gamma_\phi$ for field $\phi=S,I,R$, free energy $F$, and transmission rate $c$. The free energy $F=F_\mathrm{id}+F_\mathrm{exc}+F_\mathrm{ext}$ consists of a term $F_\mathrm{id}$ describing noninteracting persons (\ZT{ideal gas free energy}), a term $F_\mathrm{exc}$ for social interactions, i.e., social distancing and self-isolation of infected persons, and a term $F_\mathrm{ext}$ for an \ZT{external potential} describing, e.g., travel restrictions (not considered in this work). In comparison to the SIR model, social distancing is therefore incorporated explicitly, based on a microscopic model of individual persons staying away from each other. The interaction strength is measured using two parameters $C_\mathrm{sd}$ and $C_\mathrm{si}$ for social distancing and self-isolation, respectively (which are negative if the interactions are repulsive). This model is an extension of the reaction-diffusion SIR model, which has been found to give accurate predictions for the spread of the Black Death in Europe \cite{Noble1974}. Mathematical details on the SIR-DDFT model are given in the supplementary materials. DDFT is reviewed in Ref.\ \cite{teVrugtLW2020}.

As discussed in Ref.\ \cite{teVrugtBW2020}, the transmission rate $c$ should be distinguished from the effective transmission rate $c_{\mathrm{eff}}$ appearing in the standard SIR model: The former measures the transmission rate \textit{given contact}, where the amount of contacts is determined by the interactions that incorporate social distancing and self-isolation. On the other hand, the rate $c_{\mathrm{eff}}$ depends on both $c$ and the number of contacts. Consequently, it is possible in the SIR-DDFT model (but not in the SIR model) to treat these two factors separately.

This is an important advantage, since it allows to distinguish the effects of two of the main NPIs that were implemented against the COVID-19 outbreak: Face masks and other hygiene measures such as frequent hand washing reduce $c$, i.e., they decrease the probability of an infection in case of contact. Repulsive interactions, on the other hand, reduce the number of contacts. Hence, performing a parameter scan in $c$ and the interaction strength allows to distinguish the effects of the two types of measures, and thereby provides insights into the question to which extent these can supplement or replace each other.

\begin{figure}
\centering
\includegraphics{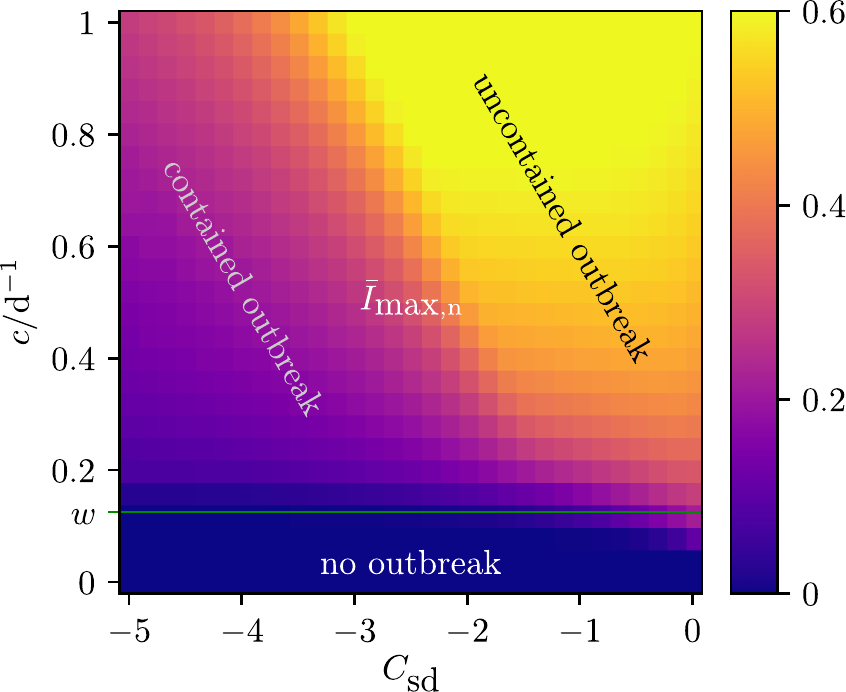}
\caption{\label{sirddftscan}\textbf{Phase diagram for the SIR-DDFT model.} The dependence of the normalized maximal number of infected persons $\bar{I}_{\mathrm{max,n}}$ on the interaction strength $C_\mathrm{sd} = \frac{1}{3} C_\mathrm{si}$ and the transmission rate $c$ is shown. Three phases (uncontained, contained, and no outbreak) are found. A reduction of contact restrictions (smaller $|C_\mathrm{sd}|$) can be compensated for by face masks/hygiene measures (smaller $c$). The green line indicates the recovery rate $w$.}
\end{figure}

To obtain the phase diagram, we have solved the SIR-DDFT model (\cref{sr,ir,rr}) numerically in two spatial dimensions with $w=0.125$/d and $m=0.0007$/d. These parameter values are adapted to the outbreak in Germany (see supplementary materials). Moreover, we set $\Gamma_S = \Gamma_I = \Gamma_R = 1$. We measure time in days (d) and everything else in dimensionless units. Population numbers shown in the plots are normalized such that the initial total population size is one. Details on the simulations can be found in the supplementary materials. The resulting phase diagram, shown in Fig.\ 1, visualizes the dependence of the normalized maximal number of infected persons $\bar{I}_{\mathrm{max,n}}$ on $c$ and on the strength of the repulsive interactions\footnote{We choose $|C_\mathrm{si}|>|C_\mathrm{sd}|$, since people will keep a larger distance from infected than from non-infected persons.} $C_\mathrm{sd} = \frac{1}{3} C_\mathrm{si}$. It is found that both a reduction of $c$ and an increase of $|C_\mathrm{sd}|$ can decrease infection numbers. The model exhibits three phases, which are characterized by low (no outbreak), intermediate (contained outbreak), and large (uncontained outbreak) infection numbers, respectively. Infection numbers are small if $c$ is below $w$ (indicated by a green line in \cref{sirddftscan}). The outbreak can be (partially) contained by large values of $|C_\mathrm{sd}|$ (even if $c$ is also large) or by intermediate values of $c$ and $C_\mathrm{sd}$. Therefore, it is possible, to a certain extent, to reduce the amount of contact restrictions (i.e., to decrease $|C_\mathrm{sd}|$) without increasing the infection numbers if the transmission rate $c$ is also reduced, which is possible by hygiene measures. Consequently, the model shows that face masks allow to re-open a society after a shutdown in a controlled way. The way in which the parameter $c$ is changed by implementing face masks depends on their efficacy and on the adherence in the population, a strong reduction of $c$ is possible if both are large (see Ref.\ \cite{HowardEtAl2020} for a quantitative estimate of the effect of face masks).

\section*{\label{long}Adaptive strategies and multiple outbreaks}
Up to now, we have assumed that the mitigation measures are imposed in the same way at all times, i.e., that the model parameters are constant. In practice, however, they will be imposed and lifted in an adaptive fashion depending on whether infection numbers rise above or fall below certain thresholds. Strategies of this form are of significant importance for the COVID-19 outbreak \cite{FergusonEtAl2020,DaviesEtAl2020}. We now discuss how such approaches can be described mathematically, starting with the simple SIRD model. 

Let us assume that a shutdown is started once the number of infected persons is larger than a threshold value $\bar{I}_\mathrm{start}$, and stopped once it falls below a value $\bar{I}_\mathrm{stop}$ with $\bar{I}_\mathrm{stop} \leq \bar{I}_\mathrm{start}$. Mathematically, this corresponds to a non-ideal relay operator (also called \ZT{rectangular hysteresis loop} or \ZT{lazy switch}), which was incorporated into the SIR model by Chladn{\'a} \textit{et al.\ }\cite{ChladnaKRR2020}. Here, we extend the model from Ref.\ \cite{ChladnaKRR2020} by also taking into account the fact that the infection rate will not jump immediately when a threshold is crossed, since a society requires some time to implement restrictions. Thus, we assume that the effective transmission rate $c_\mathrm{eff}$ converges exponentially \cite{ChowellHCFH2004} to a value $c_1$ or $c_0$ in the presence or absence of interventions, respectively. These considerations lead to the dynamical equation
\begin{equation}
\dot{c}_{\mathrm{eff}}(t) =
\begin{cases}
\alpha (c_0 - c_{\mathrm{eff}}(t)) &\text{if }(\bar{I}(\tau) < \bar{I}_\mathrm{start} \forall \tau \in [0,t])\\ 
& \text{or } (\exists t_1 \in [0,t] \text{ such that }\\ 
& \bar{I}(t_1) \leq \bar{I}_\mathrm{stop} \text{ and } 
\bar{I}(\tau) < \bar{I}_\mathrm{start} \forall \tau \in (t_1,t]),\\
\alpha (c_1 - c_{\mathrm{eff}}(t)) &\text{if }\exists t_1 \in [0,t] \text{ such that}\\ 
& \bar{I}(t_1) \geq \bar{I}_\mathrm{start} \text{ and } 
\bar{I}(\tau) > \bar{I}_\mathrm{stop} \forall \tau \in (t_1,t].
\end{cases}
\label{c}%
\end{equation}
Here, $\alpha$ is a constant parameter, and the form of \cref{c} ensures a convergence to $c_0$ or $c_1$, depending on the infection numbers and the history of the system. Usually, the initial condition will be $c_{\mathrm{eff}}(0) = c_0$, since social distancing measures are not present at the beginning of an outbreak. As discussed in the supplementary materials, realistic parameter choices for the outbreak in Germany (which we use for our simulations) are given by $\alpha = 0.206$/d, $c_0 = 0.479$/d, and $c_1 = 0.105$/d.

\begin{figure}[p]
\centering\vspace*{-12mm}
\includegraphics{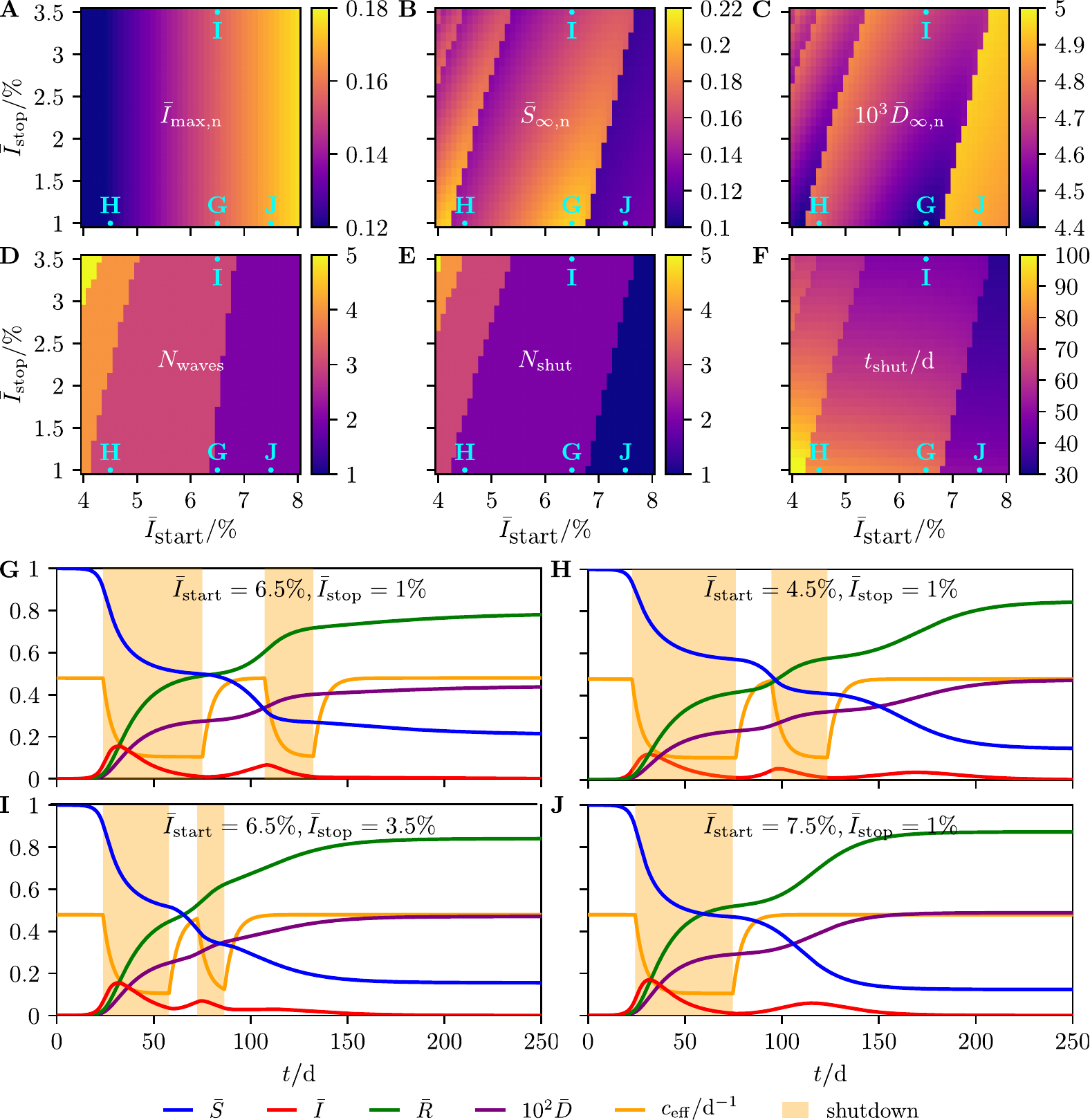}
\caption{\label{fig:adaptivestrategy}\textbf{Phase diagrams and time evolutions for the extended SIRD model with hysteresis.} The phase diagrams show \textbf{(A)} the maximal number of infected persons $\bar{I}_{\mathrm{max,n}}$, \textbf{(B)} the number of susceptibles remaining at the end of the pandemic $\bar{S}_{\infty,\mathrm{n}}$, \textbf{(C}) the final number of deaths $\bar{D}_{\infty,\mathrm{n}}$, \textbf{(D)} the number of waves $N_\mathrm{waves}$ of the pandemic, \textbf{(E)} the number of shutdowns $N_\mathrm{shut}$, and \textbf{(F)} the total shutdown time $t_\mathrm{shut}$ as a function of the shutdown thresholds $\bar{I}_\mathrm{start}$ and $\bar{I}_\mathrm{stop}$ (cf.\ \cref{c}). A variety of phases with different numbers of waves and shutdowns are observed. The time evolution of the number of susceptible $\bar{S}$, infected $\bar{I}$, recovered $\bar{R}$, and dead $\bar{D}$ persons and the effective transmission rate $c_\mathrm{eff}$ is shown for \textbf{(G)} $\bar{I}_\mathrm{start} = 6.5\%$ and $\bar{I}_\mathrm{stop} = 1\%$, \textbf{(H)} $\bar{I}_\mathrm{start} = 4.5\%$ and $\bar{I}_\mathrm{stop} = 1\%$, \textbf{(I)} $\bar{I}_\mathrm{start} = 6.5\%$ and $\bar{I}_\mathrm{stop} = 3.5\%$, and \textbf{(J)} $\bar{I}_\mathrm{start} = 7.5\%$ and $\bar{I}_\mathrm{stop} = 1\%$. Shutdown periods are indicated by shaded areas. The various parameter values lead to different numbers of outbreaks and different shutdown lengths. Blue points in Figs.\ 2A-2F indicate the parameter combinations chosen for the time simulations shown in Figs.\ 2G-2J.}
\end{figure}

The political decision that has to be made then is the choice of $\bar{I}_\mathrm{start}$ and $\bar{I}_\mathrm{stop}$, i.e., what the infection numbers should be in order for a shutdown to be started and stopped, respectively. To investigate this problem, we have solved \cref{s,i,r,c} numerically with parameter values $w=0.125$/d and $m=0.0007$/d for different values of $\bar{I}_\mathrm{start}$ and $\bar{I}_\mathrm{stop}$ in order to obtain the phase diagrams\footnote{Note that the values of $\bar{I}_\mathrm{start}$ (4-8\%) and $\bar{I}_\mathrm{stop}$ (1-3.5\%) in the phase diagrams are rather large to ensure that the behavior is more clearly visible. We have verified that different phases also exist for smaller threshold values.}. Details on the simulations can be found in the supplementary materials. We adapted the values of $w$ and $m$ to the current COVID-19 pandemic (see supplementary materials). As an initial condition, we have used the number of confirmed infections in Germany at the 10th of March 2020 (reported in Ref.\ \cite{RKI2020b} to be 1296) normalized by the population of Germany. The results are shown in Fig.\ 2, visualizing (A) the normalized maximal number of infected persons $\bar{I}_\mathrm{max,n}$, (B) the normalized number of susceptibles $\bar{S}_{\infty,\mathrm{n}}$ at the end of the pandemic (i.e., the number of persons that have never been infected), and (C) the normalized total number of deaths $\bar{D}_{\infty,\mathrm{n}}$. As can be seen from Fig.\ 2A, the maximal peak $\bar{I}_\mathrm{max,n}$ depends only on $\bar{I}_\mathrm{start}$. Hence, for avoiding a large number of infected persons at the same time and thus a collapse of the healthcare system, it is primarily important to start the shutdown sufficiently early. The point at which it is lifted again is less relevant. This observation is in agreement with results from Ref.\ \cite{FergusonEtAl2020}.

A different and much more complex result is found when considering the final number of susceptibles $\bar{S}_{\infty,\mathrm{n}}$ and the total number of deaths $\bar{D}_{\infty,\mathrm{n}}$. Here, various distinct phases can be observed, which have a staircase-shaped boundary that depends on both $\bar{I}_\mathrm{start}$ and $\bar{I}_\mathrm{stop}$. As can be expected, large values of $\bar{S}_{\infty,\mathrm{n}}$ correspond to small values of $\bar{D}_{\infty,\mathrm{n}}$ and vice versa (if fewer people are infected, fewer people die). For large values of $\bar{I}_\mathrm{start}$ (i.e., in the phase on the right), the number of deaths is large. However, within each phase, the number of deaths increases upon reducing $\bar{I}_\mathrm{start}$ at fixed $\bar{I}_\mathrm{stop}$. This is a remarkable and surprising result, since one would intuitively expect a smaller shutdown threshold to be beneficial. When reducing $\bar{I}_\mathrm{stop}$ at fixed $\bar{I}_\mathrm{start}$ within a phase, the number of deaths becomes smaller, although it jumps to a larger value if a phase boundary is crossed from above.

An explanation for the complexity of the phase diagrams can be found in Figs.\ 2D, 2E, and 2F, which show the number of waves\footnote{The number of waves is measured by the number of local maxima of the function $\bar{I}(t)$.} of the pandemic $N_\mathrm{waves}$, the number of shutdowns $N_\mathrm{shut}$, and the total shutdown time $t_\mathrm{shut}$ as a function of $\bar{I}_\mathrm{start}$ and $\bar{I}_\mathrm{stop}$. The difference between the various phases in Figs.\ 2B and 2C is the number of shutdowns $N_\mathrm{shut}$. Increasing $\bar{I}_\mathrm{stop}$ at fixed $\bar{I}_\mathrm{start}$ leads to a larger number of waves and shutdowns and a reduced total shutdown time (in agreement with Ref.\ \cite{DaviesEtAl2020}, where $\bar{I}_\mathrm{start}$ and $\bar{I}_\mathrm{stop}$ were not distinguished). However, increasing $\bar{I}_\mathrm{start}$ at fixed $\bar{I}_\mathrm{stop}$ reduces $N_\mathrm{waves}$ and $N_\mathrm{shut}$. 

Finally, a very interesting observation is that the phase boundaries for $N_\mathrm{waves}$ and $N_\mathrm{shut}$ are not at the same positions. While a larger number of shutdowns generally corresponds to a larger number of waves, reducing $\bar{I}_\mathrm{start}$ below the critical value for $n$ shutdowns (with $n \in \mathbb{N}$) does not immediately lead to $n+1$ waves because the critical value of $\bar{I}_\mathrm{start}$ for $n+1$ waves is slightly smaller. Since it is, as far as the number of deaths is concerned, beneficial to be slightly below the critical value of $\bar{I}_\mathrm{start}$ separating regions with $n$ and $n-1$ shutdowns, choosing $\bar{I}_\mathrm{start}$ in such a way that the wave $n+1$ is avoided needs careful adjustment. This requires, of course, that one is aware of the difference between the phase boundaries for $N_\mathrm{waves}$ and $N_\mathrm{shut}$, which makes our results highly relevant for political decisions on shutdown thresholds.

In Figs.\ 2G-2J, the time evolutions of $\bar{S}$, $\bar{I}$, $\bar{R}$, $\bar{D}$, and $c_\mathrm{eff}$ are shown for different combinations of $\bar{I}_\mathrm{start}$ and $\bar{I}_\mathrm{stop}$. Shutdown periods are shaded in yellow. Figure 2G, corresponding to $\bar{I}_\mathrm{start} = 6.5 \%$ and $\bar{I}_\mathrm{stop} = 1\%$, shows two shutdowns. After the first shutdown, which is longer, infection numbers rise again (\ZT{second wave}), such that a second shutdown is necessary. A third wave after the second shutdown is not observed. As can be seen from Figs.\ 2D and 2E, $\bar{I}_\mathrm{start} = 6.5 \%$ and $\bar{I}_\mathrm{stop} = 1\%$ corresponds to a choice of parameters between the phase boundaries for $N_\mathrm{waves}$ and $N_\mathrm{shut}$. In Fig.\ 2H, results for the same $\bar{I}_\mathrm{stop} = 1\%$ and smaller $\bar{I}_\mathrm{start} = 4.5\%$ are shown. Although these parameters also lead to two shutdowns, a third wave of the pandemic is observed here after the second shutdown. Therefore, the final number of deaths $\bar{D}_{\infty,\mathrm{n}}$ is larger in this case. Figure 2I shows the time evolution for $\bar{I}_\mathrm{start} = 6.5 \%$ and $\bar{I}_\mathrm{stop} = 3.5\%$, i.e., $\bar{I}_\mathrm{stop}$ is increased at fixed $\bar{I}_\mathrm{start}$ compared to Fig.\ 2G. Here, the two shutdowns are shorter and closer to each other, and $\bar{D}_{\infty,\mathrm{n}}$ is larger than in Fig.\ 2G. A third wave is also observed. Finally, Fig.\ 2J gives results for $\bar{I}_\mathrm{start} = 7.5 \%$ and $\bar{I}_\mathrm{stop} = 1\%$, i.e., $\bar{I}_\mathrm{stop}$ is the same as in Figs.\ 2G and 2H, but $\bar{I}_\mathrm{start}$ is increased into the region with $N_\mathrm{shut} = 1$. Consequently, there is only one shutdown. After it ends, a relatively large second wave occurs, leading to a relatively high overall number of deaths.

Our results have important consequences for political decisions on intervention strategies. Of course, the best strategy for keeping both $\bar{I}_{\mathrm{max,n}}$ and $\bar{D}_{\infty,\mathrm{n}}$ small is to start the shutdown early and stop it late (bottom left corner of the phase diagram). However, this is not always possible due to the social and economical costs associated with a shutdown (as can be seen in Fig.\ 2F, the total shutdown time $t_\mathrm{shut}$ is very long in this case). In practice, a political decision has to be made regarding the question when to start and end a shutdown given limited resources. 

When making a political decision on when to start and end shutdown (choosing $\bar{I}_\mathrm{start}$ and $\bar{I}_\mathrm{stop}$), one needs to take into account the existence of the various phases shown in Fig.\ 2. A small variation of the threshold values can lead to a different phase, which changes the number of outbreaks and shutdowns and thus significantly affects the total number of deaths. The optimal strategy depends on what one is aiming for: 
\begin{itemize}
    \item If the main goal is to keep $\bar{I}_\mathrm{max,n}$ small to avoid a collapse of the healthcare system, one should start the shutdown early (small $\bar{I}_\mathrm{start}$).
    \item As far as $\bar{D}_{\infty,\mathrm{n}}$ is concerned, it is beneficial to choose $\bar{I}_\mathrm{start}$ and $\bar{I}_\mathrm{stop}$ close to a phase boundary in such a way that a slight increase of $\bar{I}_\mathrm{start}$ or decrease of $\bar{I}_\mathrm{stop}$ would reduce the number of shutdowns by one.
    \item The choice of $\bar{I}_\mathrm{stop}$ also corresponds to a trade-off between $\bar{D}_{\infty,\mathrm{n}}$ and $t_\mathrm{shut}$. Increasing it within a phase at constant $\bar{I}_\mathrm{start}$ leads to a larger number of deaths and a shorter shutdown time. 
    \item Remarkably, strategies with multiple shutdowns can have advantages over strategies with a single shutdown. While in many cases more shutdowns correspond to more waves, an additional wave can be avoided after a further shutdown if the threshold values are chosen close to a phase boundary.
\end{itemize}

\section*{\label{secondwave}The \ZT{second wave} in spatiotemporal dynamics}
In practice, contact restrictions will typically be removed earlier than hygiene requirements such as face masks if infection numbers decrease. Consequently, more detailed insights can be gained using the SIR-DDFT model, in which effects of face masks and contact restrictions can (as shown in Fig.\ 1) be modeled separately. For this purpose, we introduce a dynamic equation for the interaction strength in the form 
\begin{equation}
\dot{C}_i(t) =
\begin{cases}
\alpha (C_{i,0} - C_i (t)) &\text{if }(\bar{I}(\tau) < \bar{I}_\mathrm{start} \forall \tau \in [0,t])\\ 
& \text{or } (\exists t_1 \in [0,t] \text{ such that }\\ 
& \bar{I}(t_1) \leq \bar{I}_\mathrm{stop}  \text{ and } 
\bar{I}(\tau) < \bar{I}_\mathrm{start} \forall \tau \in (t_1,t]),\\
\alpha (C_{i,1} - C_i (t)) &\text{if }\exists t_1 \in [0,t] \text{ such that}\\ 
& \bar{I}(t_1) \geq \bar{I}_\mathrm{start}\text{ and }  
\bar{I}(\tau) > \bar{I}_\mathrm{stop} \forall \tau \in (t_1,t],
\end{cases}
\label{cr}%
\end{equation}
where $i = \mathrm{sd},\mathrm{si}$. The form of \cref{cr} has been chosen in analogy to \cref{c}. Changing the interaction strength according to \cref{cr} while keeping $c$ constant models a scenario in which contact restrictions are imposed and removed depending on infection numbers while no change regarding measures such as face masks is made. Investigating the SIR-DDFT model with a dynamic interaction strength is of significant interest not only for disease spreading, but also for condensed matter physics, since it corresponds to a DDFT with a time-dependent interaction potential. Theories of this form are yet to be investigated and can therefore be expected to exhibit a variety of novel and interesting effects.

\begin{figure}
\centering
\includegraphics{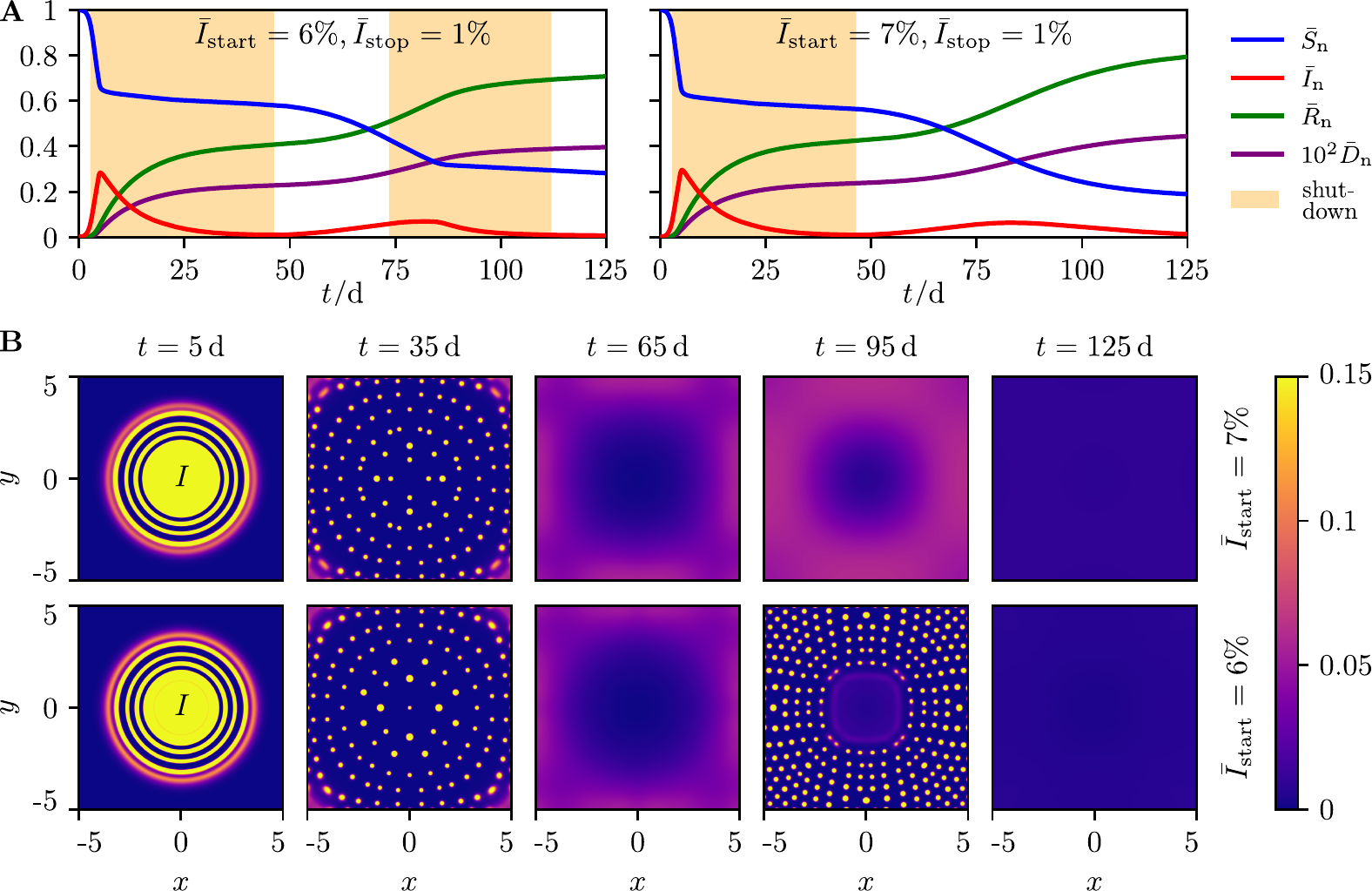}
\caption{\label{fig:3}\textbf{Multiple outbreaks in the SIR-DDFT model with hysteresis.} \textbf{(A)} Time evolution of the normalized total number of susceptible $\bar{S}_\mathrm{n}$, infected $\bar{I}_\mathrm{n}$, recovered $\bar{R}_\mathrm{n}$, and dead $\bar{D}_\mathrm{n}$ persons for $\bar{I}_\mathrm{stop}=1\%$ as well as $\bar{I}_\mathrm{start}=6\%$ (left) and $\bar{I}_\mathrm{start}=7\%$ (right). Two waves of the pandemic are observed for both parameter combinations. A second shutdown only occurs for $\bar{I}_\mathrm{start}=6\%$. \textbf{(B)} Density of infected persons $I(x,y,t)$ at different times $t$. Phase separation is observed as a consequence of repulsive interactions during a shutdown. If there are multiple shutdowns, phase separation occurs multiple times.}
\end{figure}

To study the effects of dynamic interaction strengths, we have solved \cref{sr,ir,rr,cr} numerically in two spatial dimensions with $\Gamma_S=\Gamma_I=\Gamma_R=1$, $c=0.479$/d, $w=0.125$/d, $m=0.0007$/d, $\alpha = 0.206$/d, $C_{\mathrm{sd},0}=C_{\mathrm{sd},1}=-1$, $C_{\mathrm{si},0}=-1$, and $C_{\mathrm{si},1}=-15$. Details on the simulations and the choice of parameters can be found in the supplementary materials. The results are shown in Fig.\ 3. Simulations have been performed for $\bar{I}_\mathrm{start}= 6\%$ and $\bar{I}_\mathrm{start}= 7\%$, with $\bar{I}_\mathrm{stop}=1\%$ for both cases. The resulting time evolutions, shown in Fig.\ 3A, are reminiscent of the results for the simpler model (Figs.\ 2G-2J). In both cases, a second wave of the pandemic is observed after the first shutdown. For $\bar{I}_\mathrm{start}= 6\%$, a second shutdown is initiated to inhibit the second outbreak, whereas no second shutdown is observed for $\bar{I}_\mathrm{start}= 7\%$. The simulation results thereby confirm the observations from the simpler model. However, they also add to it an important new aspect: A second wave can also occur if, after a shutdown, only contact restrictions are lifted while other measures are kept in place (constant $c$).

An effect of this type was observed in Germany: While face masks are still mandatory in public places (constant $c$), contact restrictions have been relaxed after the initial shutdown. In consequence, infection numbers have risen again \cite{BeckerEtAlH2020}. The extended SIR-DDFT model allows for a detailed investigation of a variety of shutdown strategies by adapting the values of the model parameters. In Fig.\ 3, we have chosen $c$ in such a way that it allows to recover the effective reproduction number measured in Germany in early March 2020 (corresponding to an infrequent use of face masks). The choice $C_{\mathrm{sd},0}=C_{\mathrm{si},0}=-1$ corresponds to the assumption that there is moderate social distancing in the no-shutdown phase that does not distinguish between healthy and infected persons (which can arise if infected persons cannot be easily identified as such, as it is the case for COVID-19 \cite{SaljeEtAl2020,LiPCSZYS2020}). In the case of a shutdown, a strong increase of $|C_\mathrm{si}|$ (large value of $|C_{\mathrm{si},1}|$) then reflects both an increased amount of testing (allowing for a specific isolation of infected persons) and stronger physical isolation. Other possible scenarios include a lower value of $c$ (increased use of face masks), larger values of $|C_{\mathrm{sd},0}|$ and $|C_{\mathrm{si},0}|$ (stricter social distancing in the no-shutdown phase), and larger values of $|C_{\mathrm{sd},1}|$ and $|C_{\mathrm{si},1}|$ with smaller ratio $C_{\mathrm{si},1}/C_{\mathrm{sd},1}$ (strict physical distancing in the shutdown phase without testing). Hence, the extended SIR-DDFT model is a flexible and useful tool for analyzing under which conditions and in which way a second wave will occur for a certain combination of measures. Using our freely available code \cite{CodeAndData}, simulations can be easily performed for any policy the consequences of which one wishes to investigate.

Snapshots from the time evolution of the density $I(x,y,t)$ of infected persons as a function of position $\vec{r}=(x,y)^\mathrm{T}$ are shown in Fig.\ 3B for $\bar{I}_\mathrm{start}= 6\%$ and $\bar{I}_\mathrm{start}= 7\%$. The complete time evolutions are shown in the supplementary movies S1 and S2. Initially, the infected persons are concentrated in the middle of the domain and spread outwards radially ($t = 5$\,d). Afterwards (at $t = 35$\,d), a phase separation effect is observed where the infected persons arrange into separated spots. This pattern formation, which was discussed in Ref.\ \cite{teVrugtBW2020}, can be interpreted as infected persons self-isolating at their houses. When the shutdown ends, the strength of the interactions is reduced such that phase separation is no longer present ($t = 65$\,d). For $\bar{I}_\mathrm{start}= 6\%$ (but not for $\bar{I}_\mathrm{start}= 7\%$), phase separation is observed a second time at $t=95$\,d during a second shutdown. The second phase separation differs from the first one in that it emerges from a distribution that is already rather homogeneous and not from an accumulation of infected persons in the middle. Finally, at $t = 125$\,d, there are almost no infected persons left in both simulations.

These findings are very interesting for public health policy, since they show that the first and second wave do not only differ by the initial values for $\bar{S}$, $\bar{I}$, and $\bar{R}$ -- the only aspect that can be captured in the simpler model -- but also by their different spatial distributions. This can be seen when comparing the distributions at $t=5$\,d and $t= 65$\,d, which represent the initial stages of the first and second wave, respectively. The first wave starts after a radial spread from the center, i.e., the infection is initially localized. Before the second wave, however, the disease has already spread over the entire area. This difference is also relevant for the current spread of COVID-19 in Germany: The first wave was a consequence of infected persons arriving by travel, and therefore started at isolated positions. In contrast, the second wave emerges from a more homogeneous spatial distribution \cite{Drosten2020}. From our model, this can be expected to be a common feature of second waves. Initially, a disease will always break out at single spots, which corresponds to an inhomogeneous initial condition $I(\vec{r},0)$. If contact restrictions (repulsive interactions) are lifted, the SIR-DDFT model describes a purely diffusive dynamics that typically leads to a homogeneous distribution. Therefore, the initial condition for the second wave is more homogeneous than for the first one. On the other hand, as can be seen from Fig.\ 3A, the overall infection numbers are smaller for the second wave. The snapshot for $t=95$\,d in the bottom row of Fig.\ 3B shows that phase separation does not occur at the center, where the concentration of infected persons is lower at $t= 65$\,d (initial stage of the second wave). Physically, this corresponds to a shutdown that is locally restricted as a consequence of infection numbers becoming large only in certain regions.

\section*{\label{conclusion}Discussion}
In summary, we have employed the SIR-DDFT model and an extended SIRD model with hysteresis to study the effects of different containment strategies. We have found that lifting contact restrictions can be partially compensated for by stricter hygiene measures. Investigating adaptive strategies showed that different combinations of thresholds lead to various phases. They differ by the number of waves and shutdowns and, consequently, by the number of deaths and the total shutdown time, making this effect immensely important for public health policy. Spatiotemporal simulations have revealed that a second wave can also arise if only contact restrictions are lifted, and that it tends to have a different spatial distribution than the first wave. By adapting parameter values, the model allows to study the effects of a large variety of containment strategies in any country. Possible extensions of this work include the investigation of further strategies, such as partial shutdowns or isolation of specific groups. Moreover, the SIR-DDFT model could be extended to include vaccination \cite{ElazzouziATT2019}.

\section*{Acknowledgements}
We thank Benedikt Bieringer, Markus Dertwinkel, and Julian Jeggle for helpful discussions. 
R.W.\ is funded by the Deutsche Forschungsgemeinschaft (DFG, German Research Foundation) -- WI 4170/3-1. 
The simulations for this work were performed on the computer cluster PALMA II of the University of M\"unster.

\section*{Code and data}
The code used for performing the simulations underlying this work as well as the source data for Figs.\ 1-3 and S1 are provided at Zenodo \cite{CodeAndData}.

\bibliographystyle{science}
\bibliography{refs}

\clearpage
\setcounter{page}{1}\setcounter{equation}{0}\setcounter{figure}{0}%
\renewcommand{\theequation}{S\arabic{equation}}\renewcommand{\thefigure}{S\arabic{figure}}%
\begin{center}\vspace*{3mm}{\Large\textbf{\uppercase{Supplementary Materials}}}\vspace*{7mm}\end{center}

\section*{Materials and Methods}
\subsection*{\label{math}Mathematical details of the SIR-DDFT model}
Here, we describe the SIR-DDFT model following Ref.\ \cite{teVrugtBW2020}. Dynamical density functional theory (DDFT), reviewed in Ref.\ \cite{teVrugtLW2020}, describes the time evolution of a density field $\rho\rt$. For a single-component fluid, it is given by
\begin{equation}
\partial_t\rho = \Gamma \Nabla \cdot\bigg(\rho \Nabla \frac{\delta F}{\delta \rho}\bigg)
\label{ddft}%
\end{equation}
with a mobility $\Gamma$ and a free energy $F$. Equation \eqref{ddft} can be derived from the microscopic dynamics of overdamped Brownian particles using the adiabatic approximation, which approximates the pair correlations of the nonequilibrium system by those of an equilibrium system with the same one-body density \cite{MarconiT1999}. In the case of multiple fields $\{\rho_i\}$, \cref{ddft} generalizes to
\begin{equation}
\partial_t \rho_i = \Gamma_i \Nabla \cdot \bigg(\rho_i \Nabla \frac{\delta F}{\delta \rho_i}\bigg).    
\label{ddft2}
\end{equation}
In our present work, the fields are given by $S$, $I$, and $R$ (density of susceptible, infected, and recovered persons, respectively). In addition, we need to add reaction terms to the DDFT equation \eqref{ddft2} (as done, with other physical motivations, in Refs.\cite{LutskoN2016,MonchoD2020}), since the \ZT{particles} can change their species, i.e., persons can get infected or recover. The reaction terms are obtained from the SIRD model. 
This leads to the model
\begin{align}
\partial_tS &= \Gamma_S\Nabla\cdot\bigg( S\Nabla\frac{\delta F}{\delta S}\bigg) - cSI,\label{sri}\\  \partial_tI &= \Gamma_I\Nabla\cdot\bigg( I \Nabla\frac{\delta F}{\delta I}\bigg) +cSI - wI -mI,\label{iri}\\ 
\partial_tR &=\Gamma_R\Nabla\cdot\bigg( R \Nabla\frac{\delta F}{\delta R}\bigg) + wI\label{rri}
\end{align}
with transmission rate $c$, recovery rate $w$, and death rate $m$.

The free energy $F$ has three terms:
\begin{equation}
F = F_{\mathrm{id}} + F_{\mathrm{exc}} + F_{\mathrm{ext}}.
\label{fre}%
\end{equation}
First, the ideal gas free energy
\begin{equation}
F_{\mathrm{id}} = \beta^{-1}\INT{}{}{^d r}\rho(\vec{r},t)(\ln(\rho(\vec{r},t)\Lambda^d) -1)\label{fid}
\end{equation}
describes a system of noninteracting particles with the rescaled inverse temperature $\beta$, number of spatial dimensions $d$, and thermal de Broglie wavelength $\Lambda$. In the case $F=F_{\mathrm{id}}$, \cref{ddft} simply gives the standard diffusion equation
\begin{equation}
\partial_t \rho = D\Nabla^2\rho    
\end{equation}
with $D = \Gamma \beta^{-1}$. The term $F_{\mathrm{ext}}$ describes the influence of an external potential and is set to zero throughout this work. Finally, the excess free energy $F_{\mathrm{exc}}$ describes interactions. It is not known exactly and needs to be approximated. In our case, the interactions are social interactions such as social distancing and self-isolation. The basic idea is that persons practicing social distancing can be described as repulsively interacting particles \cite{BouchnitaJ2020}. We assume that the repulsive interactions can be described by a soft (Gaussian) pair potential. The reason for this is that, even in the case of social distancing, there will still be a certain (although reduced) amount of contact. Hence, soft potentials are more appropriate than hard-core interactions. For interaction potentials as chosen here, the mean-field approximation
\begin{equation}
F_{\mathrm{exc}}=\frac{1}{2}\INT{}{}{^3r}\INT{}{}{^3r'}U_2(\vec{r}-\vec{r}')\rho(\vec{r},t)\rho(\vec{r}',t)
\end{equation}
is known to give good results \cite{ArcherE2001}. Assuming that the excess free energy $F_{\mathrm{exc}}$ contains a contribution for social distancing $F_{\mathrm{sd}}$ and a contribution for self-isolation $F_\mathrm{si}$ then gives
\begin{equation}
F_{\mathrm{exc}} = F_{\mathrm{sd}}+F_\mathrm{si}\label{free}
\end{equation}
with
\begin{gather}
\begin{split}
F_{\mathrm{sd}} = -\INT{}{}{^dr}\INT{}{}{^dr'} C_{\mathrm{sd}}e^{-\sigma_{\mathrm{sd}}(\vec{r}-\vec{r}')^2}\bigg(\frac{1}{2}S(\vec{r},t)S(\vec{r}',t)+S(\vec{r},t)R(\vec{r}',t)+\frac{1}{2}R(\vec{r},t)R(\vec{r}',t)\bigg),\label{fsd}\\
\end{split}\\
\begin{split}
F_{\mathrm{si}} = - \INT{}{}{^dr}\INT{}{}{^dr'} C_{\mathrm{si}}e^{-\sigma_{\mathrm{si}}(\vec{r}-\vec{r}')^2}I\rt\bigg(\frac{1}{2}I(\vec{r}',t) 
+ S(\vec{r}',t)+R(\vec{r}',t)\bigg)\label{fsi}.
\end{split}
\end{gather}
Here, $C_{\mathrm{sd}}$ and $C_{\mathrm{si}}$ determine the strength and $\sigma_{\mathrm{sd}}$ and $\sigma_{\mathrm{si}}$ the range of the interactions. 

Inserting \cref{fre,fid,free,fsd,fsi} into \cref{sri,iri,rri} gives the final model equations \cite{teVrugtBW2020}
\begin{align}
\partial_tS &= D_S\Nabla^2 S - \Gamma_S\Nabla\cdot\big(S \Nabla (C_{\mathrm{sd}}K_{\mathrm{sd}}\star (S+R) 
+ C_{\mathrm{si}}K_{\mathrm{si}}\star I)\big) - cS I, \label{srspec}\\  
\partial_tI &= D_I\Nabla^2 I - \Gamma_I\Nabla\cdot\big( I\Nabla(C_{\mathrm{si}}K_{\mathrm{si}}\star(S+I+R))\big)
+cS I- wI, \label{irspec}\\
\partial_tR &= D_R\Nabla^2 R - \Gamma_R\Nabla\cdot\big( R\Nabla (C_{\mathrm{sd}}K_{\mathrm{sd}}\star (S+R) 
+ C_{\mathrm{si}}K_{\mathrm{si}}\star I)\big)+ wI, \label{rrspec}
\end{align}
where $D_\phi=\Gamma_\phi \beta^{-1}$ for $\phi=S,I,R$ are the diffusion coefficients,
\begin{align}
K_\mathrm{sd}(\vec{r})&=\exp(-\sigma_\mathrm{sd}\vec{r}^2),\\
K_\mathrm{si}(\vec{r})&=\exp(-\sigma_\mathrm{si}\vec{r}^2)
\end{align}
are the kernels, and $\star$ is the spatial convolution.

\subsection*{\label{choice}Choice of parameter values}
\begin{figure}
\centering
\includegraphics{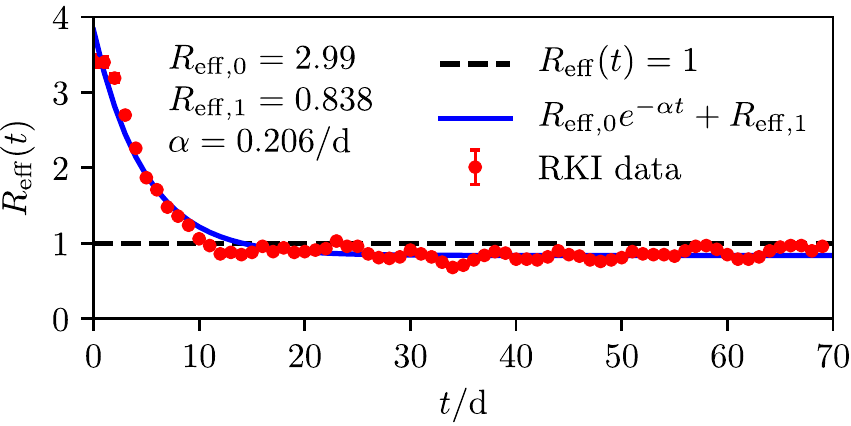}
\caption{\label{fit}Comparison of the function \eqref{function} for the time evolution of the effective reproduction number $R_\mathrm{eff}$ with data from the Robert Koch Institute \cite{RKI2020} (Interval: 03-10-2020 to 05-18-2020). The error bars of the data points are smaller than the data points. Oscillations in the empirical data arise from differences in reported case numbers between different days of the week.}
\end{figure}

The parameters for the simulations presented in the main text have been chosen in such a way that their order of magnitude is realistic for the current COVID-19 outbreak in Germany. We use days d as the unit of time and dimensionless units for all other quantities. For Germany, the effective reproduction number $R_{\mathrm{eff}}$, which in our model is given by \cite{teVrugtBW2020}
\begin{equation}
R_{\mathrm{eff}} = \frac{c_{\mathrm{eff}}\bar{S}}{w},
\label{reff}%
\end{equation}
is estimated by the Robert Koch Institute (RKI), the central public health institute of the German federal government, on a daily basis. If we assume $\bar{S} \approx N=1$ with the total population size $N$, we get 
\begin{equation}
R_\mathrm{eff}(t)=\frac{c_\mathrm{eff}(t)}{w}.    
\label{reff2}%
\end{equation}
From \cref{c} of the main text, we can infer that the approach of $R_\mathrm{eff}(t)$ to its shutdown value will be governed by a function of the form
\begin{equation}
R_{\mathrm{eff}} (t)=R_{\mathrm{eff},0}e^{-\alpha t} + R_{\mathrm{eff},1}  
\label{function}%
\end{equation}
with $R_\mathrm{eff,0}= (c_0 - c_1)/w$ and $R_{\mathrm{eff},1}= c_1/w$. 
As shown in \cref{fit}, choosing $R_{\mathrm{eff},0} = 2.99$, $R_{\mathrm{eff},1} = 0.838$, and $\alpha = 0.206$/d gives a good agreement with empirical data. Furthermore, we assume $w=0.125$/d, which is consistent with the mean infection duration of 8 days reported in Ref.\ \cite{MaierB2020}. From this, we can infer $c_0 \approx 0.479$/d and $c_1 \approx 0.105$/d.

Moreover, we assume following Ref.\ \cite{anderHeidenB2020} that the probability of dying from COVID-19 in the case of available intensive care is $p_\mathrm{d} =0.005625$. This result is given by the probability of hospitalization (0.045) multiplied by the probability of requiring intensive care given hospitalization (0.25) multiplied by the probability of dying during intensive care (0.5), which results in $p_\mathrm{d} =0.045 \times 0.25 \times 0.5 =0.005625$. Given the probability $p_\mathrm{d}$ of dying during time $T$, we can obtain the death rate $m$ (which is needed for the SIRD model) as \cite{FleurenceH2007}
\begin{equation}
m = - \frac{1}{T}\ln(1-p_\mathrm{d}). 
\label{rate}%
\end{equation}
Assuming that persons are infected for $T=8$\,d and die at a constant rate during this time (which, of course, is a strong simplification) gives $m \approx 0.0007$/d. 

For the extended SIR-DDFT model, the same parameter values for $w$, $m$, and $\alpha$ can be used. The parameter $c$ of the SIR-DDFT model is not identical to the parameter $c_{\mathrm{eff}}$ of the SIR model, which is why we discuss here how the value of $c$ can be obtained (which is important for practical applications of the SIR-DDFT model also to regions other than Germany). In the simplest case of a homogeneous distribution of the population, the relation between $c$ and $c_\mathrm{eff}$ is given by $c = c_{\mathrm{eff}} A$ with the domain area $A$ \cite{teVrugtBW2020}. In this work, we use $A=100$. If we set the total population size to $N=100$ and assume $\bar{S}\approx N$, \cref{reff} gives
\begin{equation}
R_\mathrm{eff} \approx \frac{c N}{w A} = \frac{c}{w}.
\label{reff3}
\end{equation}
Comparing \cref{reff2,reff3} shows that the value used for $c_\mathrm{eff}(0)$ can also be used for $c$ under the approximation $c_\mathrm{eff}(0)=c/A$ if the population size is set to $A$ rather than to 1. In particular, using $c \approx 0.479$/d in the spatiotemporal simulations allows to recover, in the limiting case of a  completely homogeneous distribution, the value of $R_\mathrm{eff}$ that corresponds to the values measured in Germany in early March 2020 (inserting $c = 0.479$/d into \cref{reff3} gives $R_\mathrm{eff} = 3.832$, which is approximately equal to the result $R_\mathrm{eff}(0) = 3.828$ obtained from \cref{function}). 

\subsection*{\label{numeric}Numerical analysis}
The simulations for Figs.\ 1 and 3 have been performed in two spatial dimensions on a quadratic domain $[-L/2,L/2]\times[-L/2,L/2]$ with size $L=10$ and periodic boundary conditions. We have solved the equations of the SIR-DDFT model using an explicit finite-difference scheme with spatial step size $\dif x = 0.04$ for Fig.\ 1 and $\dif x = 0.01$ for Fig.\ 3 and adaptive time steps. The shutdown state was updated explicitly every $0.01$\,d. As an initial condition, we have used a Gaussian distribution with amplitude $\approx 7.964$ and variance $L^2/50$ centered at $(x,y)=(0,0)$ for $S(x,y,0)$, $I(x,y,0)=0.001 S(x,y,0)$, and $R(x,y,0)=0$, such that the mean overall density was $1$. Regarding parameter values not specified in the main text, we have set $D_S = D_I = D_R = 0.01$ and $\sigma_{\mathrm{sd}}=\sigma_{\mathrm{si}}=100$. The simulations for Fig.\ 2 were also solved via an explicit finite-difference scheme with adaptive time steps, while the shutdown state was updated every $0.01$\,d. As an initial condition, we used $\bar{S}(0) = 1 - \bar{I}(0)$, $\bar{I}(0) = 1296/(80\times 10^6)$ (number of confirmed infections in Germany at the 10th of March 2020 \cite{RKI2020b}, normalized by the approximate population of Germany), $\bar{R}(0)=0$, $\bar{D}(0)=0$, and $c_{\mathrm{eff}}(0)=c_0$.

\end{document}